# Near-optimal single-photon sources in the solid-state

N. Somaschi[1,*], V. Giesz[1,*], L. De Santis[1,2,*], J. C. Loredo[3], M. P. Almeida[3], G. Hornecker[4,5], S. L. Portalupi[1], T. Grange[4,5], C. Anton[1], J. Demory[1], C. Gomez[1], I. Sagnes[1], N. D. Lanzillotti-Kimura[1], A. Lemaitre[1], A. Auffeves[4,5], A. G. White[3], L. Lanco[1,6] and P. Senellart[1,7,+]

1. CNRS-LPN Laboratoire de Photonique et de Nanostructures, Université Paris-Saclay, Route de Nozay, 91460 Marcoussis, France
2. Université Paris-Sud, Université Paris-Saclay, F-91405 Orsay, France
3. Centre for Engineered Quantum Systems, Centre for Quantum Computer and Communication Technology, School of Mathematics and Physics, University of Queensland, Brisbane, Queensland 4072, Australia
4. Université Grenoble Alpes, F-38000 Grenoble, France
5. CNRS, Institut Néel, "Nanophysique et Semiconducteurs" Group, F-38000 Grenoble, France
6. Département de Physique, Université Paris Diderot, 4 rue Elsa Morante, 75013 Paris, France
7. Département de Physique, Ecole Polytechnique, Université Paris-Saclay, F-91128 Palaiseau, France

* Equally contributing authors
+ corresponding author: Pascale.senellart@lpn.cnrs.fr

**Single-photons are key elements of many future quantum technologies, be it for the realisation of large-scale quantum communication networks[1] for quantum simulation of chemical and physical processes[2] or for connecting quantum memories in a quantum computer[3]. Scaling quantum technologies will thus require efficient, on-demand, sources of highly indistinguishable single-photons[4]. Semiconductor quantum dots inserted in photonic structures are ultrabright single photon sources[5–7], but the photon indistinguishability is limited by charge noise induced by nearby surfaces[8]. The current state of the art for indistinguishability are parametric down conversion single-photon sources, but they intrinsically generate multiphoton events and hence must be operated at very low brightness to maintain high single photon purity[9,10]. To date, no technology has proven to be capable of providing a source that simultaneously generates near-unity indistinguishability and pure single-photons with high brightness. Here, we report on such devices made of quantum dots in electrically controlled cavity structures. We demonstrate on-demand, bright and ultra-pure single photon generation. Application of an electrical bias on deterministically fabricated devices[11,12] is shown to fully cancel charge noise effects. Under resonant excitation, an indistinguishability of 0.9956±0.0045 is evidenced with a $g^{(2)}(0)$=0.0028±0.0012. The photon extraction of 65% and measured brightness of 0.154±0.015 make this source 20 times brighter than any source of equal quality. This new generation of sources open the way to a new level of complexity and scalability in optical quantum manipulation.**

Near future challenges in optical quantum manipulation build on the possibility of creating and manipulating a large number of single photons. The transfer rate of quantum communications scales linearly with the photon flux over short distances, and exponentially over long distances[13]. The scalability of photonic quantum computers critically depends on the photon source efficiency and quality[14,15]. Intermediate quantum computing tasks such as Boson Sampling[16–19] have an advantage that scales exponentially with increasing photon number over the best known classical strategies. In all such cases, sources should produce high-purity, highly-indistinguishable single photons, in combination with a high brightness. To date, the majority of quantum communication and quantum computing demonstrations have been performed using heralded spontaneous parametric down conversion sources (SPDC) that exhibit the best quality in terms of indistinguishability. However, such high quality is obtained only for very low source-brightness. Indeed, the photon-pair generation process leads to higher-number terms in the photon state that are detrimental to both the single-photon purity and indistinguishability[20]. This intrinsic limitation strongly restricts the number of single photons one can manipulate, with measurement times reaching typically dozens to hundreds of hours for experiments involving just 3 or 4 photons[21,22].

Strong efforts have been made in the semiconductor community to provide sources that would overcome these limitations. Fifteen years ago, semiconductor quantum dots (QDs) that are, to a large extent, artificial atoms, have been shown to emit single photons[23]. Yet, for a long time, the challenge remained to extract the photon emission from the semiconductor high refractive index material. An impressive progress has been made recently in this area, with demonstrations of very high extraction efficiencies by inserting the QD in a photonic structure like micropillar cavities[6] or nanowires on a metallic mirror[5,7]. Record brightness around 80% has been demonstrated in both systems[5,6]. Such brightness would drastically change the field of quantum photonics if the source could also meet the high standards in terms of indistinguishability set by the SPDC sources.

Single photon sources (SPSs) based on nanowires have not shown high photon indistinguishability yet, probably because of important charge noise at the wire surface[8]. For QDs in micropillar cavities, accelerating the spontaneous emission through the Purcell effect allowed the demonstration of indistinguishabilities of 80% at 40-50% brightness[6]. However such results were obtained by filling the traps around the QDs using an additional non-resonant excitation. Although efficient, this technique is not equally effective for all devices and does not allow reaching high indistinguishability at maximum brightness. Resonant fluorescence for QDs planar structures recently allowed the demonstration of close to unity indistinguishability, yet without an efficient extraction of the photons[24,25]. In gated structures, charge noise was shown to take place at low frequency, with negligible dephasing at generation frequencies higher than 50 kHz[8]. In the present work, we have implemented a control of the charge environment for QDs in connected pillar cavities. We use a fully deterministic fabrication process to apply an electric field on a cavity structure optimally coupled to a QD. Benefiting from high extraction efficiency, we demonstrate the efficient generation of pure single photon with near unity indistinguishability.

The devices are fabricated from a planar λ cavity embedding an InGaAs QD layer and surrounded by GaAs/Al$_{0.9}$Ga$_{0.1}$As distributed-Bragg-reflectors. The sample is doped to get an effective n-i-p diode structure and optimized to define a Fermi level around the QD while minimizing the free carrier losses in the mirrors. The cavity design for applying an electric field is similar to the one presented in ref.[12]: a single micropillar is connected to a surrounding circular frame by four one dimensional 1.5 µm wide wires. This frame overlaps with a large mesa where the top p-contact is defined. A standard n-contact is deposited on the back of the sample. Fig 1a presents a schematic of a single device. To have a full control of the QD-cavity coupling, we use an advanced in-situ lithography technique allowing us to position the QD within 50 nm of the pillar center and to spectrally adjust the cavity resonance to the QD transition with 0.5 nm spectral accuracy[11]. Figure 1b shows an optical microscope image of a diode where 18 sources were fabricated during the same in-situ lithography process. A photoluminescence map of one device is shown on Figure 1c: the bright QD emission in the pillar center evidences the efficient photon extraction. The fine electrical tuning of the QD exciton transition through the Stark effect is shown in figure 1d: in resonance with the cavity mode at -0.6 V, a strong enhancement of the signal is observed.

The main characteristics that define the quality of the sources are now studied, namely their purity, brightness and the indistinguishability of the successively emitted photons. The devices were first studied at 4K under a single 3-picosecond pulsed non-resonant excitation around 890 nm. We present the properties of two different QD-pillar devices (named QD1 and QD2) with a cavity quality factor Q≈12000, summarized in figures 2 and S2 (see supplementary) for pillar 1 and 2 respectively. Tuning the QD resonance to the cavity mode through the application of an electrical bias, a shortening of the radiative lifetime down to 150 ps is observed. This corresponds to a Purcell factor of $F_p$=7.6, considering a lifetime around 1.3 ns for a QD exciton in bulk. Under these conditions, the single photon purity is characterized in a standard Hanbury Brown and Twiss setup. Figure 2a shows a typical curve, evidencing a high single photon purity with an autocorrelation function at zero delay $g^{(2)}(0)$= 0.024±0.007.

The brightness of the source, defined as the number of photons collected per excitation pulse into the first lens, is given by (β $\eta_{out}$ $p_x$) with β=$F_p$/($F_p$+1) the fraction of emission into the mode, $\eta_{out}$=$\kappa_{top}$/$\kappa$ the out-coupling efficiency defined as the ratio between the photon escape rate through the top of the cavity to the total escape rate, and $p_x$ the occupation factor of the QD state. To measure the source brightness the overall setup efficiency is characterized. As shown in figure 2c, the source brightness increases with power since $p_x$ scales as (1-exp(-P/$P_{sat}$)) where $P_{sat}$ is the saturation power of the transition. From the measured count rate on the detector at maximum power, we derive a brightness value of 0.65±0.07 for QD1, consistent with $p_x$=1, β=0.88 and $\eta_{out}$=0.70 as measured through reflectivity measurements (see supplementary)[6].

To study the indistinguishability of the photons successively emitted by the device, the SPS is excited twice every laser pulse cycle (12.2 ns) with a delay of 3 ns. The successively emitted photons are temporally overlapped using a fiber beam splitter and a delay line and sent to a free space Hong-Ou-Mandel interferometer[26]. The outputs of the interferometer are coupled to two single photon detectors to measure the photon correlation events.

A typical photon correlation histogram is presented in Fig.2 b. The highly reduced intensity of the zero delay peak with respect to ±3 ns peaks is the direct signature of a high coalescence probability. Figure 2c presents the overall device characterization as a function of excitation power, where the mean wave packet overlap of the photons M is deduced following ref.[27]. At a measured brightness of 0.65±0.07, the indistinguishability reaches M=0.78±0.07 (M=0.74±0.07) with (without) correction from the non-zero measured $g^{(2)}(0)$=0.024±0.007. As opposed to ref [6], the maximum indistinguishability is obtained here at maximum brightness. Similar results are also obtained on QD 2 shown in figure S2, giving a strong indication of a reduced charge noise influence in these electrically controlled sources.

Considering that the devices operate in the strong Purcell regime, the observed indistinguishability actually reaches the theoretical limit under non-resonant excitation. Indeed, under non-resonant pumping, the relaxation time of carriers to the lowest QD state introduces a time uncertainty on the exciton creation time that becomes comparable to the exciton radiative recombination time itself. Kiraz et al.[28] predicted that the photon indistinguishabiliy should be limited to 70-80% for the Purcell enhancement reported here, a limit reached in these measurements. Such consideration support the assumption that charge noise is actually efficiently cancelled in these gated cavity structures and that fully indistinguishable photons should then be produced under strictly resonant excitation.

To test this hypothesis, the devices are studied under strictly resonant excitation. To do so, shaped laser pulses, with temporal width of 15 ps were coupled from the top of the device, directly through the cavity mode. In this sample, the neutral exciton states show fine structure splitting (FSS) in the 10-15 µeV range. Because of the 1D wires connecting the pillar to the frame, the cavity also presents a small polarization splitting in the 90µeV range. The laser polarization is set along one direction of the cavity axes (named V), which is roughly oriented 45° with respect to the QD axes. A V polarization thus creates a coherent superposition of both exciton states that temporally evolves toward the orthogonal state coupled to the H polarized cavity mode. This evolution happens with a time scale inversely proportional to the FFS. The emission is collected in polarization H, orthogonal to the excitation polarization (figure 3.a.), with an extinction ratio of the scattered laser light of around $10^5$. The ratio between the scattered laser light and the fluorescence signal of few percent is further reduced using an etalon filter with 10 µeV (7 pm) bandwidth.

The optical characterization of two sources is presented in figures 3 and S1 for QD3 and QD4 respectively. A near-to-zero value of $g^{(2)}(0)$=0.0028±0.0012 is evidenced in figure 3b. Using a highly optimized fibered HOM interferometer, the indistinguishability of successively emitted photons is measured. From the vanishing count at zero delay on the histograms of Fig.3 **c** we extract M=0.9956±0.0045 (0.989±0.004) with (without) correction for the $g^{(2)}(0)$, revealing an almost perfect two photon coalescence. As a test, the experiment was repeated preparing the two incoming photons in cross polarization. Under this fully distinguishable condition, Fig.3 d shows a vanishing interference with an extracted M=0.057±0.084.

For completeness, Fig.3 **e** presents $g^{(2)}(0)$ and indistinguishability M as a function of the excitation power normalized to $P_\pi$, the excitation power corresponding to a π-pulse. Very high indistinguishability values above 0.973 are observed on the full power range. Calibration of the setup allows using the photon count rate into the number of polarized photons collected at the first lens, yielding a brightness of 0.16±0.02. In this device as in QD2, the photon extraction ($\beta\, \eta_{out}$) is around 0.63. The reduced brightness as compared to the case of non-resonant excitation comes from the detection in crossed polarization: the polarization rotation process is limited by the exciton FSS and the strong Purcell effect in the V mode. With a FSS of 15 μeV and the Purcell factor of $F_P$=9.8, the occupation factor of the H-emitting exciton $p_X$ reaches 0.23 at π-pulse. Implementing a side excitation, for instance taking advantage of the lateral 1D wires to guide the excitation, would bring the source brightness to 0.65 for the same device, keeping all other characteristics unchanged.

At this stage, a comparison with other QDs and SPDC sources is needed to fully evaluate the potential of the devices presented in this work. The sources are fully characterized by three figures of merit: the purity (technically, the second-order correlation function, $g^{(2)}(0)$), the indistinguishability (M), and the brightness. Since 3D diagrams are rather inconvenient, we plot in Figure 4 the source indistinguishability as a function of brightness, considering experimental results where only marginal $g^{(2)}(0) < 5\%$ were reported. To ensure a fair comparison between different devices, the plotted M value is not corrected from the $g^{(2)}(0)$. The top axis of the graph corresponds to a deterministic SPS while the right axis corresponds to a purely indistinguishable SPS. The ideal source stands in the upper right corner of the diagram.

Previous state-of-the-art results in QD systems[6,24,29] are indicated as open symbols. Since a SPS should provide photons with a well-defined polarization, the source brightness is divided by a factor of 2 for unpolarised SPS like the one presented in figure 2 or references[6,29] for non-resonant excitation. No correction is needed for QD sources based on resonant fluorescence in crossed polarization geometry.

Measured experimental data acquired for a pulsed SPDC source are also presented with grey symbols. The SPDC brightness is defined as the mean photon-number per spatial-mode (see Supplemental Material), while assuming perfect collection and detection efficiency. This corresponds to the notion of brightness at the first lens in solid-state devices. Such method allows for a fair comparison of different types of sources, independently of the driving repetition rate and the photon bandwidth. Note that the interfering photons belong to the same down-conversion event, in which case near-unity indistinguishabilities can be observed. Interference between photons from independent sources---where independent heralding for each photon can be achieved--- are typically limited to considerably lower values[20]. The experimental points presented in figure 4 here are limited to values where $g^{(2)}(0)<0.05$, corresponding to a brightness below 0.01. Indeed, the single photon purity of an SPDC source rapidly degrades with the source brightness, reaching $g^{(2)}(0)=0.25$ for a brightness of 0.07. The dotted line in figure 4 shows the theoretical limit one can expect from a SPDC source considering the corresponding experimental parameters for losses and detector efficiencies (see supplementary for details on the SPDC measurements).

Figure 4 is clear evidence that our solid-state sources bring the technology of single photon generation to a new level. Under strictly resonant excitation, the brightness of

QD sources is enhanced by a factor 20 as compared to state of the art SPDC sources. The photon purity and indistinguishability reaches ultimate values of $g^{(2)}(0)$= 0.0028±0.0012 and M=0.9956±0.0045, a quality that is perfectly adapted for highly demanding applications like fault tolerant linear optical quantum computation.

A unique characteristic of our devices is that they were obtained with a fully deterministic technology, allowing reproducible and thus scalable device fabrication. Under non-resonant excitation, a brightness as large as 0.65 is demonstrated with an indistinguishability of 0.78. Such device is highly suited for Boson sampling experiments with a large number of photons[30]. Using resonant excitation, our sources are of the highest quality, and more than an order-of-magnitude brighter than currently used SPDC sources. Since data rates decrease exponentially with the number of photons involved, such technology promises to spectacularly change the experimental landscape in quantum photonics.

**Acknowledgments:** This work was partially supported by the ERC Starting Grant No. 277885 QD-CQED, the French Agence Nationale pour la Recherche (grant ANR QDOM) the French RENATECH network, the Labex NanoSaclay, and the EU FP7 Grant No. 618072 (WASPS), by the Centre for Engineered Quantum Systems (Grant No. CE110001013) and the Centre for Quantum Computation and Communication Technology (Grant No. CE110001027). J. C. L., M. P. A. and A. G. W. thank M. Ringbauer and M. Goggin for insightful discussions, and thank the team from the Austrian Institute of Technology for kindly providing the time-tagging modules for the SPDC measurements. The LPN-CNRS authors are very thankful to Anna Nowak for her help on the technology. N.D.L.K. was supported by the FP7 Marie Curie Fellowship OMSiQuD. M. P. A. acknowledges support from the Australian Research Council Discovery Early Career Awards (No. DE120101899). A. G. W. was supported by the University of Queensland Vice- Chancellor's Senior Research Fellowship.

**Author contributions:**
Optical measurements on the QD devices were conducted primarily by V.G. N.S. and L.d.S., with help from L.L., S.L.P. and P.S. The electrically controlled samples were fabricated by N.S. with help from C.A. The sample was grown by C.G. and A.L., and the etching performed by I.S. The measurements on the SPDC sources and analysis of that data were conducted by J.L. and M.P.A, with help from A.G.W. Theoretical support to the experiment was provided by G.H., T.G. and A.A. The project was conducted by P.S. with help from L.L. All authors discussed the results and participated to manuscript preparation.

**Author information:** Correspondence and requests for materials should be addressed to Pascale.Senellart@lpn.cnrs.fr

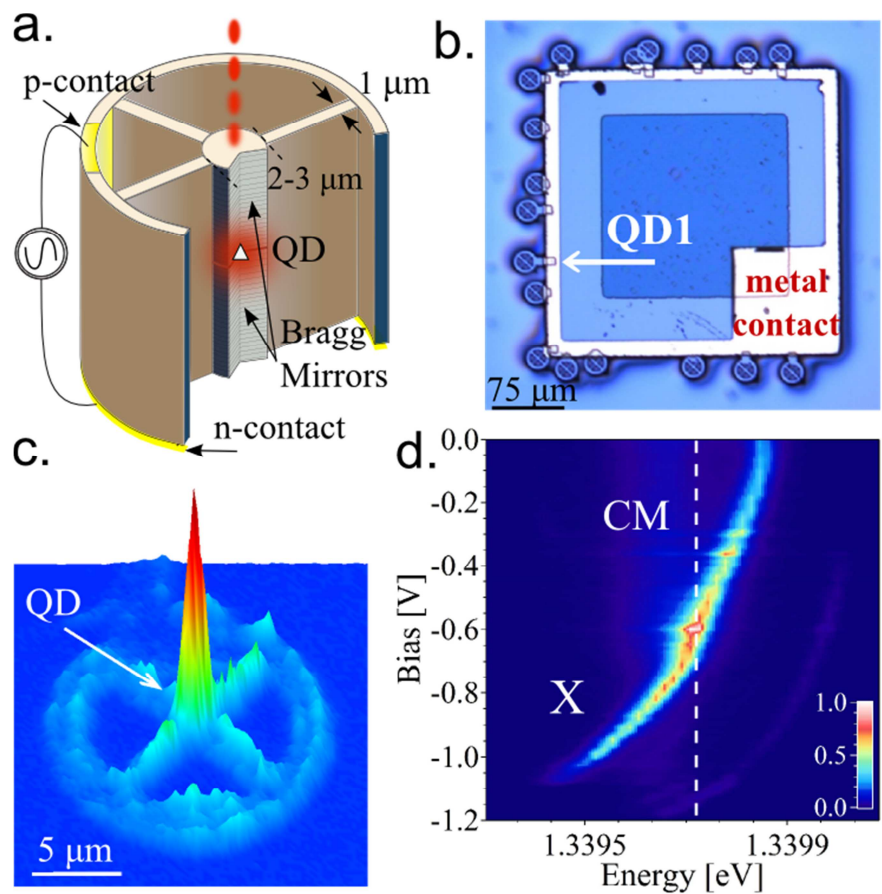

**Fig.1: Electrically controlled single photon sources: a:** Schematic of the devices under study: a micropillar coupled to a QD is connected to a surrounding circular frame by four one-dimensional wires. The top p-contact is defined on a large mesa adjacent to the frame. The n-contact is deposited on the back of the sample. **b:** Optical microscope image showing 18 connected pillar sources electrically controlled through the metallic contact defined on the 300x300 µm² diode. **c:** Photoluminescence map of a connected device: the bright emission at the centre of the device arises from the deterministically coupled QD. **d:** Emission intensity as a function of bias and energy, showing the Stark tuning of the exciton transition (X) within the cavity mode (CM) resonance (dotted line).

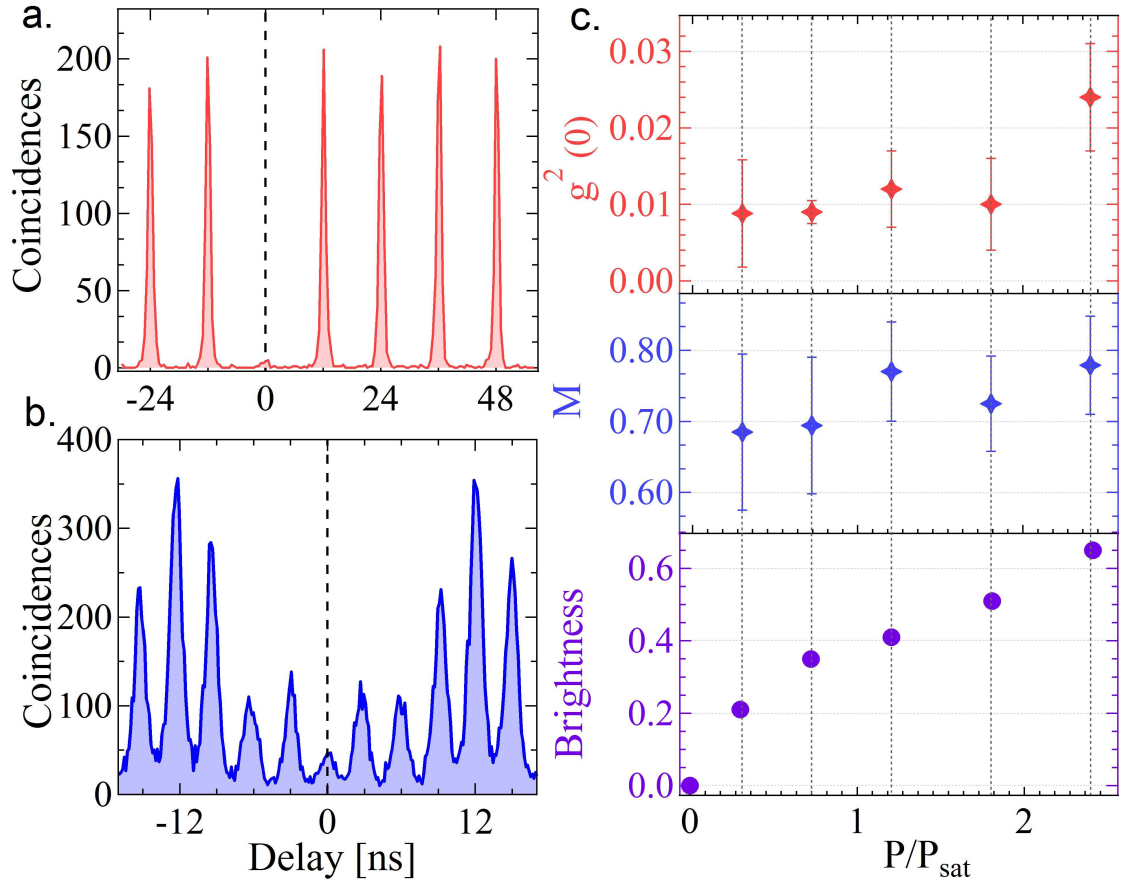

**Fig.2: Characteristics of single photon source QD1 under non-resonant excitation.**
**a:** Second-order autocorrelation histogram of device QD1 at 2.45$P_{sat}$ showing a pure single photon emission with $g^{(2)}(0)$= 0.024±0.007. b: Correlation histogram measuring the indistinguishability of photons successively emitted by QD1 (with acquisition time of 8 min). **c :** Summary of the source properties as a function of excitation power : from top to bottom: purity ($g^{(2)}(0)$) , indistinguishability (M) and brightness (collected photon per pulse).

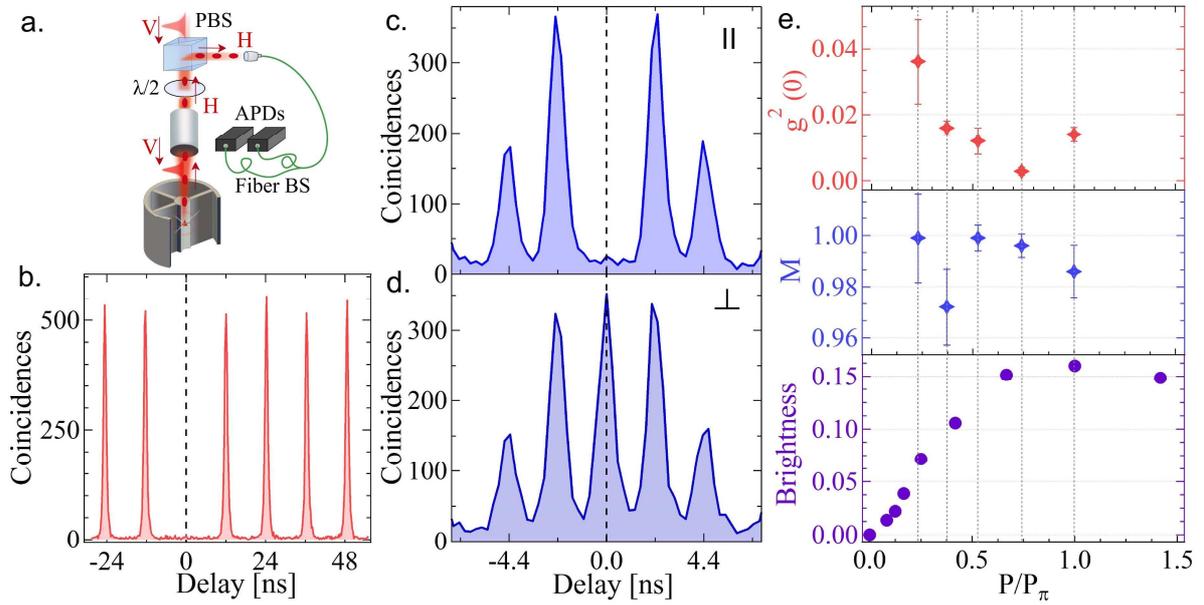

**Fig. 3: Characteristics of single photon source QD3 under resonant excitation**. **a.** Schematic of the cross polarization excitation/detection setup implemented for resonant fluorescence measurements and single photon statistics analysis. Temporally shaped laser pulses are sent from the top of the pillars and focused using a microscope objective. The emission is collected through the same objective in a confocal geometry. A polarizing beam splitter (PBS) and half wave plate allows separating crossed polarized emission from the excitation. **b.** Second-order autocorrelation histogram of device QD3 at $0.75P_\pi$ showing a pure single photon emission with $g^{(2)}(0)= 0.0028\pm0.0012$. **c-d**: Correlation histogram measuring the indistinguishability of photons successively emitted by the QD3. The photons are sent to the Hong-Ou-Mandel beamsplitter with (c) the same polarization or (d) orthogonal polarization. The acquisition time is 10 min for each curve. **e :** Summary of the source properties as a function of excitation power : from top to bottom: purity ($g^{(2)}(0)$), indistinguishability (M) and brightness (collected photon per pulse).

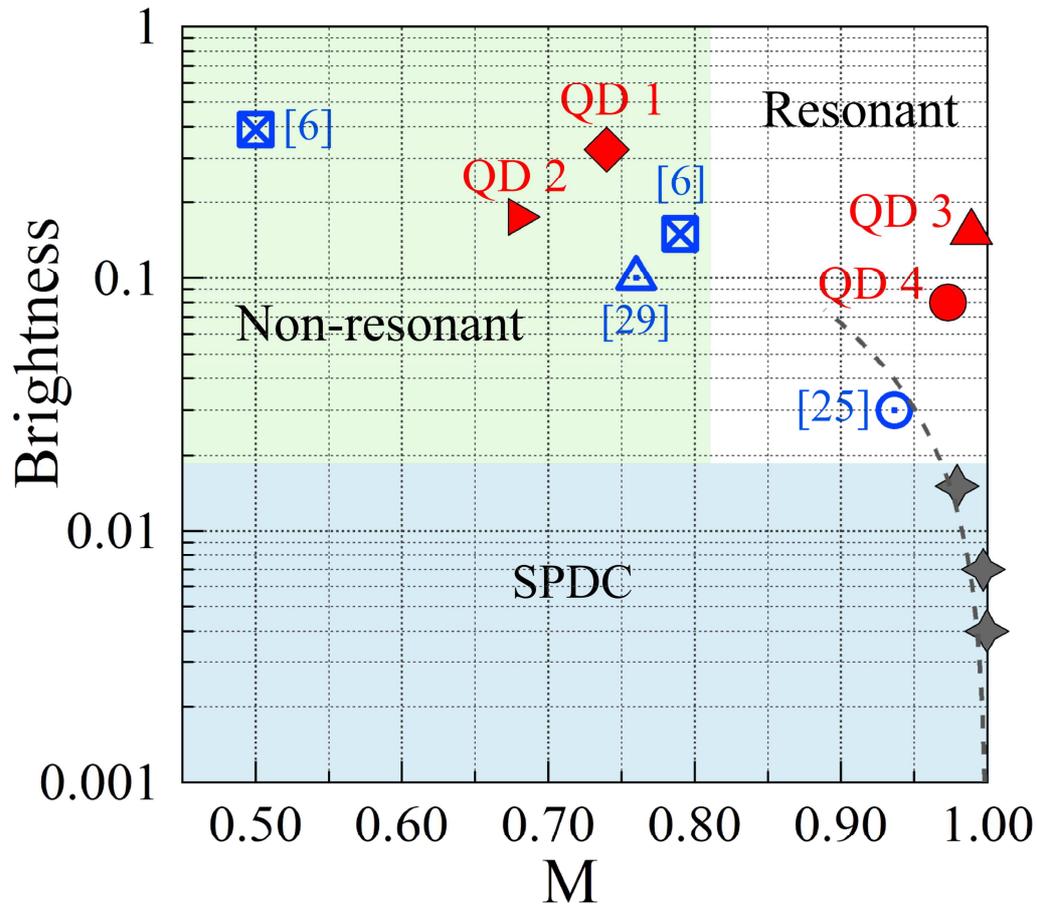

**Fig. 4: Comparison with other QD and SPDC single-photon sources**. Comparison of state of the art QD based single photon sources (blue open symbols, the corresponding reference is indicating in the label), high quality SPDC heralded single photon sources (grey symbols) and the devices reported in the present work (red symbols). QD1 and QD2 correspond to the measurements under non-resonant excitation presented in figure 2 and S2 respectively. QD3 and QD4 correspond to measurements under resonant excitation shown in figure 3 and S1 respectively. See text for a detailed discussion.

# Supplemental Material of
# "Near-optimal single photon sources in the solid-state"

N. Somaschi*,[1] V. Giesz*,[1] L. De Santis*,[1,2] J. C. Loredo,[3] M. P. Almeida,[3] G. Hornecker,[4] S. L. Portalupi,[1] T. Grange,[4] C. Anton,[1] J. Demory,[1] C. Gomez,[1] I. Sagnes,[1] N. D. Lanzillotti-Kimura,[1] A. Lemaitre,[1] A. Auffeves,[4] A. G. White,[3] L. Lanco,[1,5] and P. Senellart[1,6]

[1] CNRS-LPN Laboratoire de Photonique et de Nanostructures,
Universit Paris-Saclay, Route de Nozay, 91460 Marcoussis, France
[2] Universit Paris-Sud, Universit Paris-Saclay, F-91405 Orsay, France
[3] Centre for Engineered Quantum Systems,
Centre for Quantum Computer and Communication Technology,
School of Mathematics and Physics, University of Queensland,
Brisbane, Queensland 4072, Australia
[4] CEA/CNRS/UJF joint team Nanophysics and Semiconductors,
Institut Nel-CNRS, BP 166, 25 rue des Martyrs, 38042 Grenoble Cedex 9, France
[5] Dpartement de Physique, Universit Paris Diderot,
4 rue Elsa Morante, 75013 Paris, France
[6] Dpartement de Physique, Ecole Polytechnique,
Universit Paris-Saclay, F-91128 Palaiseau, France

## QD MEASUREMENTS

### Experimental setups

The experimental setup is based on a confocal geometry where the same microscope objective (NA 0.75) serves simultaneously for quantum dot (QD) excitation and photoluminescence (PL) emission collection. The excitation is provided by a tunable Ti-Sapph laser, providing 3 ps pulses at a 82 MHz repetition rate. For HOM measurements, each excitation pulse is split into two equally intense pulses separated by 3ns (2.2$ns$) delay for non-resonant (resonant) measurements. The sample is kept at 4 Kelvin in a close-cycle cryostat with an exchange gas. The collected PL signal is spatially diverted from the laser path through a beam splitter prior to being coupled into a single mode optical fiber.

For non resonant excitation experiments, the excitation energy is set to 1.3381eV (1.3396eV) for QD1(QD2). The signal is sent to a free space HOM interferometer: the two photon interference takes place on a non-polarizing beam splitter with $R = 0.45$ and $T = 0.50$. The signal at each output of the beam splitter is sent to a monocromator coupled to a single-photon avalanche diode (SPAD). The overall detection efficiency of the setup is estimated to be 0.25%. Considering the maximal count rate measured on the SPAD at saturation (0.125 MHz for QD1 and 0.068 MHz for QD2), the brightness was estimated to be $65 \pm 6\%$ for QD1 and $35 \pm 3\%$ for QD2.

For resonant excitation experiments, the laser energy is set to the V polarized cavity mode energy. The pulses are sent to a pulse shaper in order to obtain 15 ps pulses, corresponding to an optimal overlap with the cavity decay rate. The crossed polarization fluorescence signal, filtered with an etalon with 10 $\mu$eV bandwidth (transmission 70%) is sent to a fiber based HOM interferometer: the two photon interference takes place in a $R = 0.508$, $T = 0.492$ fibered beam splitter. The output signals are directly sent to fibered SPADs. In this configuration, the setup efficiency was measured to be 2.9%. Considering the maximal count rate measured on the SPAD at $\pi$ pulse (0.38 MHz for QD3 and 0.19 MHz for QD4), the brightness was estimated to be 16% and 8% for QD3 and QD4.

**Analysis of photon statistics.**

The values for the meanwavepacket overlap $M$ and the second-order autocorrelation function at zero delay ($g^2(0)$) were extracted from the correlation histograms of events at the output of the Hong-Ou-Mandel and Hanbury Brown and Twiss setups. For all measurements, the correlation curves are fitted with multiple peaks with a double-exponential decay shape. SPADs dark counts around $100-1000$ counts/sec lead to a small time independent background. The fit includes a constant baseline to account for these dark counts contribution. The area of the peaks are used to extract $M$ and $g^2(0)$.

The mean-wave packet overlap M is deduced using:

$$M = \frac{1}{(1-\epsilon)^2}\left[2g^{(2)}(0) + \frac{R^2+T^2}{2RT} - \frac{A_0}{A_{-2.2\text{ns}}+A_{+2.2\text{ns}}}\left(2 + g^{(2)}(0)\frac{(R^2+T^2)}{RT}\right)\right] \quad \text{(S1)}$$

where $(1-\epsilon)$ is the classic visibility of the interferometer, 0.95 (0.9988) for non-resonant (resonant) excitation. The quantities $A_0$ and $A_{-1}$ ($A_{+1}$) define the area of the peak at 0 delay and at $-1$ ($+1$) unity of delay.

Fig.S1 presents the correlation histograms (not corrected from any background) with related fit for all the four devices characterized in the present work. Panels **a** and **b** refer to QD1 and QD2, both investigated under non-resonant excitation at pumping power corresponding to the maximal source brightness. In case of QD1 we report an indistinguishability value M corrected (not-corrected) from the measured non-zero $g^2(0) = 0.024 \pm 0.007$, of $M_1 = 0.78 \pm 0.07$ ($M = 0.74 \pm 0.07$) while QD2 presents a value of $g^2(0) = 0.047 \pm 0.009$ to which corresponds a corrected visibility $M_2 = 0.77 \pm 0.08$ (not-corrected: $M = 0.68 \pm 0.081$).

Panel **c** and **d** of figure S1 present the two-photon interference histograms and fit for the photons emitted by QD3 and QD4 in resonance fluorescence. With measured near-to-zero $g^2(0) = 0.0028 \pm 0.0012$, QD3 presents an indistinguishability of $M_3 = 0.9945 \pm 0.0045$ ($0.989 \pm 0.004$) corrected (not corrected) for the $g^2(0)$. For QD4, we measure $M_4 = 0.979 \pm 0.026$ ($0.973 \pm 0.026$) corrected (not corrected) for the $g^2(0) = 0.0035 \pm 0.0040$. The results of QD3 refers to a pulse power of $0.75 P_\pi$ while the ones of QD4 to a power $P = P_\pi$.

The error on the extracted $M$ and $g^2(0)$ are calculated from the Poisson noise of the measured signal coming from fluctuations of the excitation power as well as errors on the measured parameters $R$, $T$, $(1-\epsilon)$.

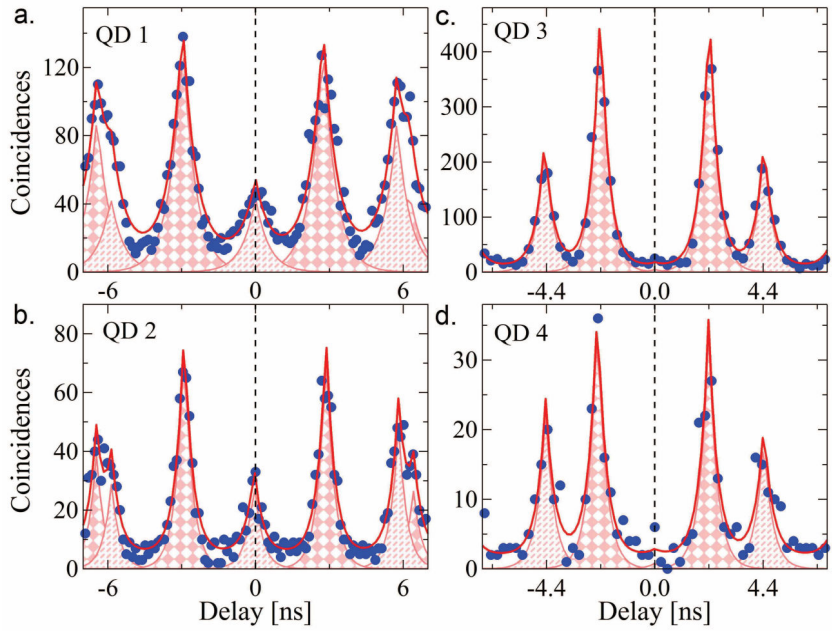

FIG. S1. **Indistinguishability under non-resonant and resonant excitation.** Correlation histograms and relative fit for all the devices studied and discussed in the present work: QD1 (**a**) and QD2 (**b**) investigated under non-resonant excitation, QD3 (**c**) and QD4 (**d**) studied under resonant pumping.

**Characterization of pillar 2 under non-resonant excitation**

We present here the characteristics of a second device (QD2) under non-resonant excitation. The exciton lifetime is $\tau_{QD} = 180$ ps corresponding to a Purcell factor of $F_p = 6.2$. Examples of $g^2(0)$ and $M$ measurements are presented in figure S2.a and b. All characteristics measured as a function of power are summarized in S2.c.

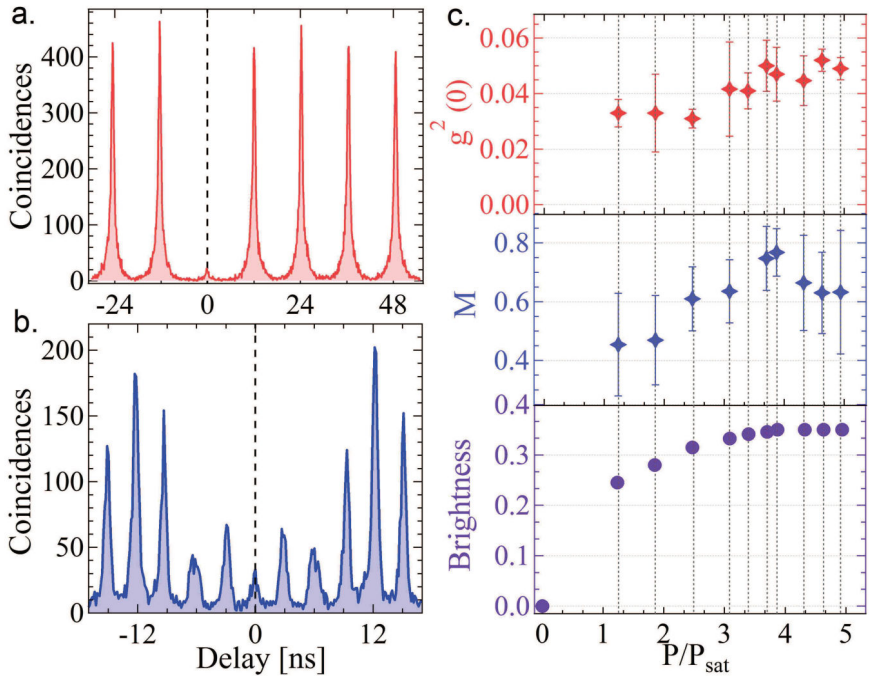

FIG. S2. **Characteristics of single photon source QD2 under non-resonant excitation.** Autocorrelation histogram of device QD2 at $3.9 P_{sat}$ showing **a.** single photon emission purity of $g^2(0) = 0.047 \pm 0.009$ and **b.** two-photon indistinguishability of $M = 0.77 \pm 0.081$ (with acquisition times of 11 min). **c.** Summary of the source properties as a function of applied laser power: from bottom to top, brightness (collected photon per pulse), indistinguishability (M) and purity ($g^2(0)$).

**Cavity characteristics of device QD3.**

Fig.S3 presents the measured reflectivity of the H polarized cavity mode for the device QD3, when the exciton is detuned from the mode. A reflectivity dip is observed corre-sponding to the cavity resonance. The $FWHM$ of the reflectivity dip gives the total cavity damping, $\kappa = 120 eV$, corresponding to a Q-factor of $Q = 11100$.

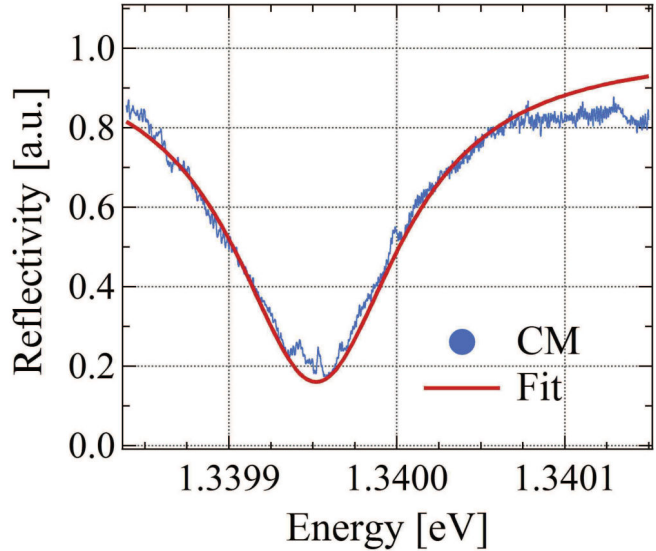

FIG. S3. **Reflectivity.** Reflectivity spectra and relative fitting of device QD3 measured at high excitation power.

The minimum reflectivity measured at the cavity resonance is given by

$$R_{\min} = \left(1 - \frac{2\kappa_{\text{top}}}{\kappa}\right)^2$$

From this, we deduce the out-coupling efficiency of $\eta_{out} = \frac{\kappa_{\text{top}}}{\kappa} = 0.7 \pm 0.05$, i.e. a measure of the light coupling through the top mirror

## SPDC MEASUREMENTS.

Our source consists of photon-pairs emission from spontaneous parametric down-conversion (SPDC) in a *beta*-barium borate (BBO) crystal pumped by a frequency-doubled mode-locked femtosecond Ti:Sapphire laser operating at 76 MHz. The state at the output of the down-conversion process is a two-mode squeezed state, and it can be written as [*Rev. Mod. Phys.* **79**, 135-174 (2007)]:

$$|\Psi_{SPDC}\rangle = \sqrt{1 - |\lambda|^2} \sum_{n=0}^{\infty} \lambda^n |n, n\rangle, \tag{S2}$$

where $\lambda$ is the squeezing parameter (being $|\lambda|^2$ proportional to the laser pump power), and $|n\rangle$ is the $n$-photon Fock state. Thus, the probability of creating $n$ photon pairs is simply given by $p(n)=(1-|\lambda|^2)|\lambda|^{2n}$.

Here, it is useful to notice that $p(n+1)/p(n)=|\lambda|^2$. That is, the ratio between the probability of creating $n+1$ photon pairs to that of $n$ pairs is determined and increases monotonically with $|\lambda|$. Therefore, although mostly consisting of vacuum, if one wishes to operate $|\Psi_{SPDC}\rangle$ as a heralded single-photon source $|1,1\rangle$, where the detection of one photon flags the presence of its *twin* photon, then we must run the source at low pump powers to achieve $|\lambda| \ll 1$, so the probability of creating $|2,2\rangle$ states (or more higher-order terms) in Eq. (S2) is negligible as compared to the only non-zero state of interest $|1,1\rangle$. These higher-order terms are responsible of degrading the visibility of two-photon interference experiments and decreasing the performance of quantum information protocols [arXiv:0808.0794].

However, obviously, keeping $|\lambda|$ too small will importantly reduce the available count rates in experiments. Thus, one must find a compromise in as how large $|\lambda|$ can be to provide decent event rates while simultaneously being small enough to minimise the impact of higher-order terms. Moreover, it turns out that these terms are more likely to survive setup losses and contribute to accumulated statistics. This can be seen from considering a simple model for optical losses: losses in one spatial mode are assumed to be the result of tracing out the reflecting port of a beam-splitter with transmittance $t$. It can be shown that the probability of loss-survival for the term $|n\rangle$ is $p_t(n)=1-(1-t)^n$, see Fig. (S4). Limited detector efficiency can also be modeled as optical loss followed by detection with unity efficiency.

From this, it is clear that a pump-dependent analysis of the source must be carried out to quantify all these effects. Indeed, we have performed such analysis by measuring two-photon interference visibilities to quantify photon indistinguishability, and the second-order autocorrelation function $g^2(0)$ at zero delay quantifying source purity.

First, we must parameterise $|\lambda|$ in relation to the used pump powers. This can be done with the lowest powers available, where detected rates of singles and coincidence counts reveal the specific value of $|\lambda|^2$ used. From here, $|\lambda|^2$ will simply be proportional to the power. Furthermore, for a comparison with brightness in solid-state sources, we use the average photon number per mode $\mu=\langle \hat{n} \otimes \hat{I}\rangle=\langle \hat{I} \otimes \hat{n}\rangle=|\lambda|^2/(1-|\lambda|^2)$ as the brightness for SPDC sources. This is a reasonable brightness parameter as it represents, in the limit

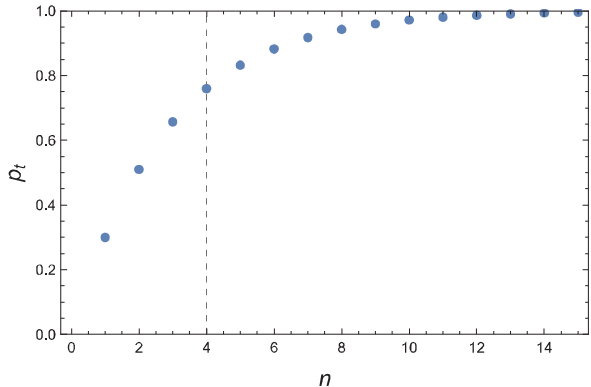

FIG. S4. Probability of the term $|n\rangle$ passing through losses in an experimental setup with transmittance $t=0.3$. The dashed line indicates a region (left) with $n<4$, containing stronger terms contributing to $|\Psi_{SPDC}\rangle$.

$|\lambda| \ll 1$, the probability per laser-pulse of one down-converted event reaching the first setup lens, and it accounts for small contributions from higher-order events intrinsic to this sources.

It is known that for a perfectly balanced 50:50 beam-splitter and completely indistinguishable photons, coincidence measurements in two-photon interference experiments will fully vanish. However, the general case of partially-distinguishable photons evolving through linear optical elements results in coincidence probabilities given by [arXiv:1403.3433]:

$$c = \frac{(1+M)}{2} |\mathrm{per}(L)|^2 + \frac{(1-M)}{2} |\det(L)|^2, \tag{S3}$$

where $M$ is the degree of indistinguishability, and $\mathrm{perm}(L)$ ($\det(L)$) is the permanent (determinant) of the matrix $L$ characterising the involved linear transformation. From Eq. (S3), the observed two-photon interference visibility $v=1-c/c_0$, with $c_0$ the coincidence probability for $M=0$, relates to $M$ via:

$$M = \left(\frac{D+P}{D-P}\right) v, \tag{S4}$$

with $D=|\det(L)|^2$, and $P=|\mathrm{per}(L)|^2$.

In our experiment, we employed a $1/3 : 2/3$ beam-splitter in a free-space configuration with single-mode input/output spatial modes. A characterisation of the corresponding linear matrix $L$ results in:

$$L = \begin{pmatrix} \sqrt{0.3310} & \sqrt{0.6690} \\ \sqrt{0.6632} & -\sqrt{0.3368} \end{pmatrix}, \tag{S5}$$

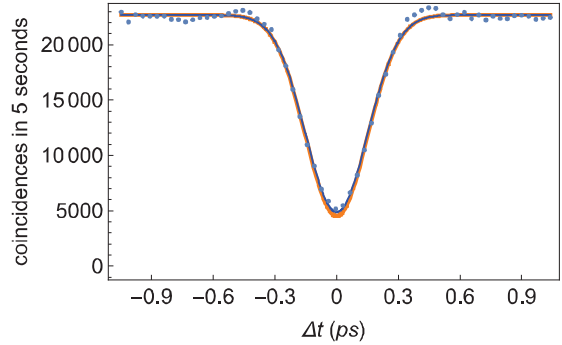

FIG. S5. Two-photon interference as a function of the temporal overlap between the interfering photons. Blue curve is a Gaussian fit with a visibility $v=(78.48 \pm 0.04)\%$. Orange curve describes the same Gaussian fit, but with a visibility equal to $(D-P)/(D+P)=80.12\%$, corresponding to the case $M=1$, with $D$ and $P$ calculated from characterisation of $L$. From these considerations, an indistinguishability value of $M=(97.95 \pm 0.05)\%$ is extracted.

where the elements $|L_{ij}|=\sqrt{t_{ij}}$ are determined from the normalised transmissions $t_{ij}$ of the $i-th$ input to the $j-th$ output. Thus, the maximum two-photon interference visibility that can be observed in such a setup is $(D-P)/(D+P)=80.12\%$. Figure S5 shows our measurement of two-photon interference for $\mu=0.015$, from which $M=(97.95 \pm 0.05)\%$ is extracted.

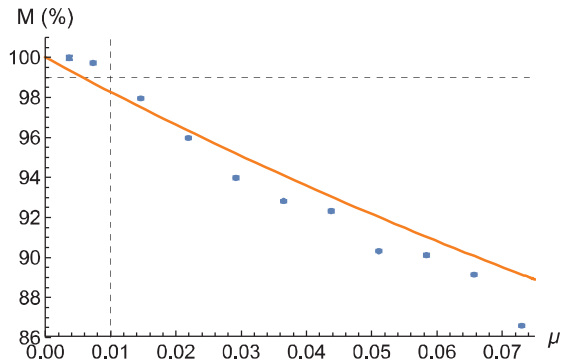

FIG. S6. Measured indistinguishability as a function of brightness. Orange line is a theoretical prediction [*New. J. Phys.* **17**, 043030 (2015)] that only depends on losses and detector efficiencies. Dashed lines indicate that $M>99\%$ can only be reached for a brightness $\mu<0.01$.

In Fig. S6, we show our measurements of $M$ as a function of the brightness $\mu$. The expected impact of optical losses, limited detector efficiency, and higher-order terms [*New. J. Phys.* **17**, 043030 (2015)] agrees well with our measured data. The model used has no free parameters and only depends on losses and detector efficiencies measured in our experimental setup.

To complete the quantitative analysis of our source, we also performed $g^2(0)$ measurements [*Europhys. Lett.* **1**, 173 (1986)] as a function of brightness. Our results are summarised in Fig. S7.

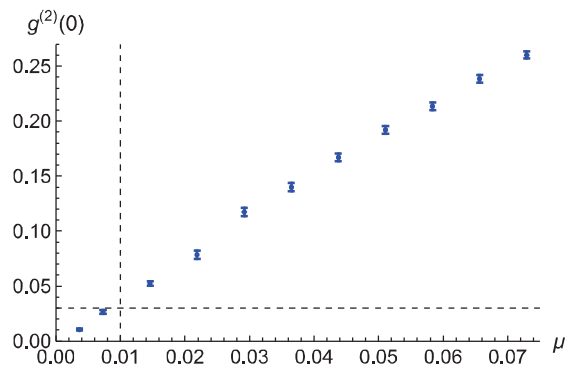

FIG. S7. Source purity and brightness. Dashed lines indicate that $g^2(0)<0.03$ is achieved only for a brightness $\mu<0.01$.

Multiplexing schemes [*Nat. Comms.* **4**, 2582] that can boost SPDC performance have not been considered for a fair comparison between the intrinsic behaviour of SPDC sources and our reported solid-state devices. From Figs. S6, and S7, it can be concluded that in order to operate a SPDC source at its highest-quality, namely $M>99\%$ and $g^2(0)<0.03$, simultaneously, the source must be operated at a brightness as low as $\mu<0.01$.

## SINGLE PHOTON PURITY AND BRIGHTNESS

For completeness, we present in Figure S8 the single photon purity $g^2(0)$ as a function of brightness for the data presented in Fig.4 of the main manuscript.

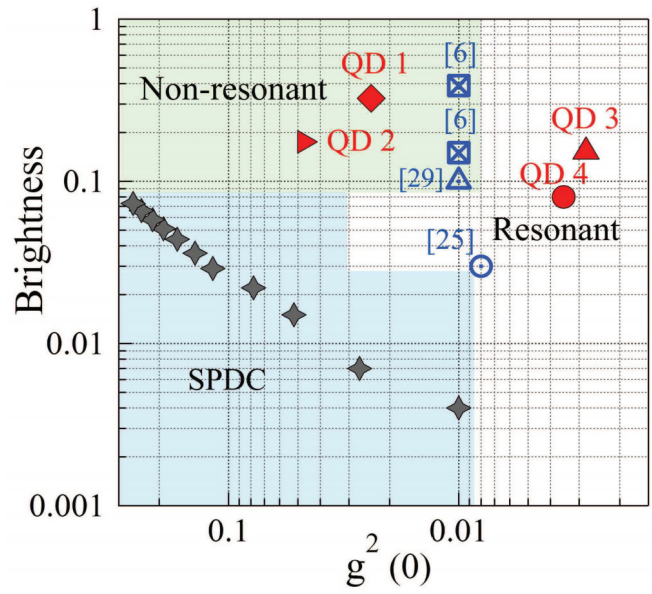

FIG. S8. **Brightness as a function of the single photon purity.** This figure is complementary of figure 4 of the main manuscript where we plot the brightness as a function of $M$ for the same experimental data. The best quality corresponds to the top right corner.